\documentclass[a4paper, 11pt]{article}
\usepackage{amsmath}
\usepackage{graphicx}
\usepackage{latexsym}

\def\b{\begin{eqnarray}}
\def\e{\end{eqnarray}}
\def\n{\noindent}

\begin{document}

\begin{center}

{\LARGE\textbf{Poisson structure and Action-Angle variables for the
Camassa-Holm equation
\\}} \vspace {10mm} \vspace{1mm} \noindent

{\large \bf Adrian Constantin$^\dag$} and {\large \bf Rossen
Ivanov$^\ast$}\footnote{On leave from the Institute for Nuclear
Research and Nuclear Energy, Bulgarian Academy of Sciences, Sofia,
Bulgaria.} \vskip1cm \n \hskip-.3cm
\begin{tabular}{c}
\hskip-1cm $\phantom{R^R}${\it School of Mathematics, Trinity
College Dublin,}
\\ {\it Dublin 2, Ireland} \\ {\it Tel:  + 353 - 1 - 608 2898 }\\{\it  Fax:  + 353 - 1- 608 2282} \\
\\{\it $^\dag$e-mail: adrian@maths.tcd.ie} \\
\\ {\it $^\ast$e-mail: ivanovr@maths.tcd.ie} \\
\\
\hskip-.8cm
\end{tabular}
\vskip1cm
\end{center}

\vskip1cm

\begin{abstract}
\n The Poisson brackets for the scattering data of the Camassa-Holm
equation are computed. Consequently, the action-angle variables are
expressed in terms of the scattering data.



\end{abstract}

\newpage

\section{Introduction}
\label{intro} The Camassa-Holm equation (CH)
\begin{equation}\label{eq1}
 u_{t}-u_{xxt}+2\omega u_{x}+3uu_{x}-2u_{x}u_{xx}-uu_{xxx}=0,
\end{equation}
where $\omega$ is a real constant, firstly appeared in \cite{FF81}
as an equation with a bi-Hamiltonian structure. In \cite{CH93} it
was pushed forward as a model, describing the unidirectional
propagation of shallow water waves over a flat bottom, see also
\cite{J02}. CH is a completely integrable equation
\cite{BBS98,CM99,C,C01,L02,R02}, describing permanent and breaking
waves \cite{CE98,M,C00}. Its solitary waves are stable solitons if
$\omega > 0$ \cite{BBS99,CS02,J03} or peakons if $\omega = 0$
\cite{CS00}. CH arises also as an equation of the geodesic flow for
the $H^{1}$ right-invariant metric on the Bott-Virasoro group (if
$\omega > 0$) \cite{M98,CKKT04}  and on the diffeomorphism group (if
$\omega = 0$) \cite{CK02,CK03}. The bi-Hamiltonian form of
(\ref{eq1}) is \cite{CH93,FF81}:

\begin{equation}\label{eq2}
 m_{t}=-(\partial-\partial^{3})\frac{\delta H_{2}[m]}{\delta m}=-(2\omega \partial +m\partial+\partial m)\frac{\delta H_{1}[m]}{\delta m}.
\end{equation}

\n where \b\label{eq4a} m = u-u_{xx} \e \n and the Hamiltonians are
\b \label{eq2a} H_{1}[m]&=&\frac{1}{2}\int m u dx
 \\\label{eq2b}
H_{2}[m]&=&\frac{1}{2}\int(u^{3}+uu_{x}^{2}+2\omega u^{2})dx. \e

\n The integration is from $-\infty$ to $\infty$ in the case of
Schwartz class functions, and over one period in the periodic case.

In general, there exists an infinite sequence of conservation laws
(multi-Hamiltonian structure) $H_n[m]$, $n=0,\pm1, \pm2,\ldots$,
including (\ref{eq2a}) and (\ref{eq2b}), such that \cite{L05}
\begin{equation}\label{eq2aa}
 (\partial-\partial^{3})\frac{\delta H_{n}[m]}{\delta m}=(2\omega \partial +m\partial+\partial m)\frac{\delta H_{n-1}[m]}{\delta m}.
\end{equation}

The CH equation can be written as

\begin{equation}\label{eq1a}
 m_{t}=\{m, H_{1}\},
\end{equation}

\n where the Poisson bracket is defined as

\begin{equation}\label{PB}
 \{A,B\}\equiv \int \frac{\delta A}{\delta m}(-2\omega \partial -m\partial-\partial m)\frac{\delta B}{\delta
 m}dx,
\end{equation}

\n or in more obvious antisymmetric form

\begin{equation}\label{PBa}
 \{A,B\}=-\int (\omega+m)\Big(\frac{\delta A}{\delta m}\partial \frac{\delta B}{\delta
 m}-\frac{\delta B}{\delta m}\partial \frac{\delta A}{\delta
 m}\Big)dx.
\end{equation}

CH has an infinite number of conserved quantities. Schemes for the
computation of the conservation laws can be found in
\cite{FS99,R02,L05,CL05,I05}.

The equation (\ref{eq1}) admits a Lax pair \cite{CH93,C01}

\b \label{eq3} \Psi_{xx}&=&\Big(\frac{1}{4}+\lambda
(m+\omega)\Big)\Psi
 \\\label{eq4}
\Psi_{t}&=&\Big(\frac{1}{2\lambda}-u\Big)\Psi_{x}+\frac{u_{x}}{2}\Psi+\gamma\Psi
\e

\n where $\gamma$ is an arbitrary constant.  We will use this
freedom for a proper normalization of the eigenfunctions.

We consider the case where $m$ is a Schwartz class function,
$\omega>0$ and $m(x,0)+\omega > 0$ (see \cite{C,CM99} for a
discussion of the periodic case).
Then $m(x,t)+\omega > 0$ for all $t$ \cite{C01}.  Let
$k^{2}=-\frac{1}{4}-\lambda \omega$, i.e.

\b \label{lambda} \lambda(k)= -\frac{1}{\omega}\Big(
k^{2}+\frac{1}{4}\Big).\e

The spectrum of the problem (\ref{eq3}) under these conditions is
described in \cite{C01}. The continuous spectrum in terms of $k$
corresponds to $k$ -- real. The discrete spectrum (in the upper half
plane) consists of finitely many points $k_{n}=i\kappa _{n}$,
$n=1,\ldots,N$ where $\kappa_{n}$ is real and $0<\kappa_{n}<1/2$.

For all real $k\neq 0$ a basis in the space of solutions of
(\ref{eq3}) can be introduced, fixed by its asymptotic when
$x\rightarrow\infty$  \cite{C01} (see also \cite{ZMNP}):

\b \label{eq5} \psi_{1}(x,k)&=&e^{-ikx}+o(1), \qquad
x\rightarrow\infty;
 \\\label{eq6}
\psi_{2}(x,k)&=& e^{ikx}+o(1), \qquad x\rightarrow \infty. \e

\n   Another basis can be introduced, fixed by its asymptotic when
$x\rightarrow -\infty$:

\b \label{eq5a} \varphi_{1}(x,k)&=&e^{-ikx}+o(1), \qquad
x\rightarrow -\infty;
 \\\label{eq6a}
\varphi_{2}(x,k)&=& e^{ikx}+o(1), \qquad x\rightarrow -\infty. \e

\n For all real $k\neq 0$ if $\Psi(x,k)$ is  a solution of
(\ref{eq3}), then $\Psi(x,-k)$ is also a solution, thus

\b \label{eq5aa} \varphi_{1}(x,k)=\varphi_{2}(x, -k), \qquad
\psi_{1}(x,k)=\psi_{2}(x, -k).
 \e

\n Due to the reality of $m$ in (\ref{eq3}) for any $k$ we have

\b \label{eq6aa} \varphi_{1}(x,k)=\bar{\varphi}_{2}(x, \bar{k}),
\qquad \psi_{1}(x,k)=\bar{\psi}_{2}(x, \bar{k})
 \e

The vectors of each of the bases are a linear combination of the
vectors of the other basis:
 \b \label{eq7}
\varphi_{i}(x,k)=\sum_{l=1,2}T_{il}(k)\psi_{l}(x,k) \e

\n where the matrix  $T(k)$ defined above is called the scattering
matrix. For real $k\neq 0$, instead of $\varphi_{1}(x,k)$,
$\varphi_{2}(x,k)$, $\psi_{1}(x,k)$, $\psi_{2}(x,k)$ due to
(\ref{eq6aa}), for simplicity we can write correspondingly
$\varphi(x,k)$, $\bar{\varphi}(x,k)$, $\psi(x,k)$,
$\bar{\psi}(x,k)$. Thus $T(k)$ has the form

\b \label{T} T(k) = \left( \begin{array}{cc}
   a(k)&  b(k)  \\
  \bar{b}(k) &  \bar{a}(k) \\
\end{array}  \right) \,
\e

\n and clearly

\b \label{eq8} \varphi(x,k)=a(k)\psi(x,k)+b(k)\bar{\psi}(x,k). \e

\n The Wronskian $W(f_{1},f_{2})\equiv
f_{1}\partial_{x}f_{2}-f_{2}\partial_{x}f_{1}$ of any pair of
solutions of (\ref{eq3}) does not depend on $x$. Therefore

\b \label{eq9} W(\varphi(x,k), \bar{\varphi}(x,k))= W(\psi(x,k),
\bar{\psi}(x,k))=2ik \e

\n From (\ref{eq8}) and (\ref{eq9}) it follows that

\b \label{eq10} |a(k)|^{2}-|b(k)|^{2}=1, \e

\n i.e. $\det (T(k))=1$. Computing the Wronskians
$W(\varphi,\bar{\psi})$ and $W(\psi,\varphi)$ and using (\ref{eq8}),
(\ref{eq9}) we obtain:

\b \label{eq10a}a(k)&=&(2ik)^{-1}
\Big(\bar{\psi}_{x}(x,k)\varphi(x,k)-\bar{\psi}(x,k)\varphi_{x}(x,k)\Big),
 \\\label{eq10b}
b(k)&=&(2ik)^{-1}
\Big(\varphi_{x}(x,k)\psi(x,k)-\varphi(x,k)\psi_{x}(x,k)\Big).\e

In analogy with the spectral problem for the KdV equation, the
quantities $\mathcal{T}(k)=a^{-1}(k)$ and $\mathcal{R}(k)=b(k)/a(k)$
represent the transmission and reflection coefficients respectively.
Indeed, the asymptotic of the eigenfunction $\varphi(x,k)/a(k)$ when
$x\rightarrow\infty$ is

\b \label{eq11}
\frac{\varphi(x,k)}{a(k)}=e^{-ikx}+\mathcal{R}(k)e^{ikx}+o(1),\e

\n i.e. a superposition of incident ($e^{-ikx}$) and reflected
($\mathcal{R}(k)e^{ikx}$) waves. For $x\rightarrow-\infty$ we have a
transmitted wave:

\b \label{eq12}
\frac{\varphi(x,k)}{a(k)}=\mathcal{T}(k)e^{-ikx}+o(1)\e

\n From (\ref{eq10}) it follows that the scattering matrix is
unitary, i.e.

\b \label{eq13} |\mathcal{T}(k)|^{2}+|\mathcal{R}(k)|^{2}=1. \e

\n In what follows we will show that the entire information about
$T(k)$  in (\ref{T}) is provided by $\mathcal{R}(k)$ for $k>0$ only.
It is sufficient to know $\mathcal{R}(k)$ only on the half line
$k>0$, since from (\ref{eq5aa}) and (\ref{eq8}), $\bar{a}(k)=a(-k)$,
$\bar{b}(k)=b(-k)$ and thus $\mathcal{R}(-k)=\bar{\mathcal{R}}(k)$.
Also, from (\ref{eq13})

\b \label{eq13a} |a(k)|^{2}=(1-|\mathcal{R}(k)|^{2})^{-1}, \e

\n i.e. $|\mathcal{R}(k)|$ determines $|a(k)|$. In the next section
we will show  that $|a(k)|$ uniquely determines $\arg(a(k))$ as
well.

At the points of the discrete spectrum, $a(k)$ has simple zeroes
\cite{C01}, therefore the Wronskian $W(\varphi,\bar{\psi})$
(\ref{eq10a}) is zero. Thus $\varphi$ and $\bar{\psi}$ are linearly
dependent:

\b \label{eq200} \varphi(x,i\kappa_n)=b_n\bar{\psi}(x,-i\kappa_n).\e

\n  In other words, the discrete spectrum is simple, there is only
one (real) eigenfunction $\varphi^{(n)}(x)$, corresponding to each
eigenvalue $i\kappa_n$, and we can take this eigenfunction to be

\b \label{eq201}\varphi^{(n)}(x)\equiv \varphi(x,i\kappa_n)\e

\n Moreover, one can argue (see \cite{ZMNP}), that (cf.
(\ref{eq200}), (\ref{eq6aa}) and (\ref{eq8}))

\b \label{eq202} b_n= b(i\kappa_n)\e

\n The asymptotic of $\varphi^{(n)}$, according to (\ref{eq5a}),
(\ref{eq6}), (\ref{eq200}) is

\b \label{eq203} \varphi^{(n)}(x)&=&e^{\kappa_n x}+o(e^{\kappa_n
x}), \qquad x\rightarrow -\infty;
 \\\label{eq204}
\varphi^{(n)}(x)&=& b_n e^{-\kappa_n x}+o(e^{-\kappa_n x}), \qquad
x\rightarrow \infty. \e

\n The sign of $b_n$ obviously depends on the number of the zeroes
of $\varphi^{(n)}$. Suppose that
$0<\kappa_{1}<\kappa_{2}<\ldots<\kappa_{N}<1/2$. Then from the
oscillation theorem for the Sturm-Liouville problem \cite{B},
$\varphi^{(n)}$ has exactly $n-1$ zeroes. Therefore

\b \label{eq205} b_n= (-1)^{n-1}|b_n|.\e

The set

\b \label{eq206} \mathcal{S}\equiv\{ \mathcal{R}(k)\quad (k>0),\quad
\kappa_n,\quad |b_n|,\quad n=1,\ldots N\} \e

\n is called scattering data. In what follows we will compute the
Poisson brackets for the scattering data and we will also express
the Hamiltonians for the CH equation in terms of the scattering
data. The derivation is similar to that for other integrable
systems, e.g. \cite{ZMNP,ZF71,ZM74,FT87,BFT86}.

The time evolution of the scattering data can be easily obtained as
follows. From (\ref{eq8}) with $x\rightarrow\infty$ one has

\b \label{eq14} \varphi(x,k)=a(k)e^{-ikx}+b(k)e^{ikx}+o(1). \e The
substitution of $\varphi(x,k)$ into (\ref{eq4}) with
$x\rightarrow\infty$ gives

\b \label{eq15} \varphi_{t}=\frac{1}{2\lambda}\varphi_{x}+\gamma
\varphi \e

\n From (\ref{eq14}), (\ref{eq15}) with the choice
$\gamma=ik/2\lambda$ we obtain

\b \label{eq16} \dot{a}(k,t)&=&0,
 \\\label{eq17}
\dot{b}(k,t)&=& \frac{i k}{\lambda }b(k,t), \e

\n where the dot stands for derivative with respect to $t$. Thus
 \b \label{eq18} a(k,t)=a(k,0), \qquad b(k,t)=b(k,0)e^{\frac{i
k}{\lambda }t}; \e

\b \label{eq19} \mathcal{T}(k,t)=\mathcal{T}(k,0), \qquad
\mathcal{R}(k,t)=\mathcal{R}(k,0)e^{\frac{i k}{\lambda }t}. \e

In other words, $a(k)$ is independent on $t$ and will serve as a
generating function of the conservation laws.

The time evolution of the data on the discrete spectrum is found as
follows. $i\kappa_n$ are zeroes of $a(k)$, which does not depend on
$t$, and therefore $\dot {\kappa}_n =0$. From (\ref{eq202}) and
(\ref{eq17}) one can obtain
 \b \label{eq207} \dot{b}_n=\frac{4\omega
\kappa_n}{1-4\kappa_n^2}b_n. \e

The conservation laws are expressed through the scattering data in
Section \ref{int} and the Poisson brackets for the scattering data
are computed in Section \ref{PB}.
\section{Conservation laws and scattering data} \label{int}

The solution of (\ref{eq3}) can be represented in the form
\begin{equation}\label{eqi1}
 \varphi(x,k)=\exp \Big( -ikx + \int _{-\infty}^{x}\chi(y,k)dy
 \Big).
\end{equation}

\n For $\mathrm{Im}\phantom{*} k>0$ and $x\rightarrow \infty$,
$\varphi(x,k)e^{ ikx}=a(k)$, i.e.

\begin{equation}\label{eqi2}
 \ln a(k)= \int _{-\infty}^{\infty}\chi(x,k)dx, \qquad \mathrm{Im}\phantom{*} k>0.
\end{equation}

\n Since $a(k)$ does not depend on $t$, the expressions $\int
_{-\infty}^{\infty}\chi(x,k)dx$ represent integrals of motion for
all $k$. The equation for $\chi(x,k)$ follows from (\ref{eq3}) and
(\ref{eqi1})

\begin{equation}\label{eqi3}
 \chi_x (x,k)+\chi^2-2ik\chi=-\frac{1}{\omega}\Big(k^2+\frac{1}{4}\Big)m(x)
\end{equation}

\n and admits a solution with the asymptotic expansion
\begin{equation}\label{eqi4}
 \chi(x,k)= p_1 k+p_0+\sum_{n=1}^{\infty}\frac{p_{-n}}{k^n}.
\end{equation}

\n The substitution of (\ref{eqi4}) into (\ref{eqi3}) gives the
following quadratic equation for $p_1$:

\begin{equation}\label{eqi5}
 p_1 ^{2} -2ip_1+\frac{m}{\omega}=0,
\end{equation}

\n with solutions

\begin{equation}\label{eqi6}
 p_1=i\Big(1\pm \sqrt{1+\frac{m}{\omega}}\Big)
\end{equation}

\n Since $\int _{-\infty}^{\infty}p_1(x)dx$ is an integral of the CH
equation, presumably finite, we take the minus sign in (\ref{eqi6}).
One can easily see that $p_0$ and all $p_{-2n}$ are total
derivatives \cite{I05} and thus we have the expansion

\begin{equation}\label{eqi7}
 \ln a(k)= -i\alpha k+\sum_{n=1}^{\infty}\frac{I_{-n}}{k^n},
\end{equation}

\n where $\alpha$ is a positive constant (integral of motion):

\begin{equation}\label{eqi8}
 \alpha= \int _{-\infty}^{\infty}\Big(\sqrt{1+\frac{m(x)}{\omega}}-1\Big)dx,
\end{equation}

\n and $I_{-n}=\int _{-\infty}^{\infty}p_{-n}dx$ are the other
integrals, whose densities, $p_{-n}$ can be obtained reccurently
from (\ref{eqi3}), (\ref{eqi4}) \cite{I05}. For example

\begin{equation}\label{eqi8a}
 p_0= \frac{q_x}{4q},\qquad q\equiv m+\omega,
\end{equation}

\begin{equation}\label{eqi8aa}
 p_{-1}= \frac{1}{8}p_1+i\frac{\sqrt{\omega}}{8}\Big[\frac{1}{\sqrt{q}}-\frac{1}{\sqrt{\omega}}
 +\frac{q_{x}^{2}}{4q^{5/2}}+\Big(\frac{q_x}{q^{3/2}}\Big)_x\Big],
\end{equation}

\n etc., i.e.
\begin{equation}\label{eqi8aaa}
 I_{-1}= -\frac{1}{8}i\alpha+i\frac{\sqrt{\omega}}{8}\int _{-\infty}^{\infty}\Big(\frac{1}{\sqrt{q}}-\frac{1}{\sqrt{\omega}}
 +\frac{q_{x}^{2}}{4q^{5/2}}\Big)dx, \qquad \ldots
\end{equation}

The asymptotic of $a(k)$ for $\mathrm{Im}\phantom{*} k>0$ and
$|k|\rightarrow\infty$ from (\ref{eqi7}) is $a(k)\rightarrow
e^{-i\alpha k}$, or

\begin{equation}\label{eqi9}
 e^{i\alpha k}a(k)\rightarrow 1, \qquad \mathrm{Im}\phantom{*} k>0,
 \qquad |k|\rightarrow\infty.
\end{equation}

Now let us consider the function

\begin{equation}\label{eqi10}
 a_1(k)\equiv e^{i\alpha k}\prod _{n=1}^{N}\frac{k+i\kappa_n}{k-i\kappa_n}a(k).
\end{equation}

\n This function is analytic for $\mathrm{Im}\phantom{*} k>0$, but
does not have any zeroes there. This is due to the fact \cite{C01}
that $a(k)$ has at most simple zeroes at the points of the discrete
spectrum $i\kappa_n$. Therefore $\ln a_1 (k)$ is analytic in the
upper half plane and due to (\ref{eqi9}) $\ln a_1 (k)\rightarrow 0$
for $|k|\rightarrow\infty$. Moreover, on the real line
$|a_1(k)|=|a(k)|$, and the Kramers-Kronig dispersion relation
\cite{J99} for the function

\begin{equation}\label{eqi11}
 \ln a_1(k)=\ln |a(k)|+i \arg a_1(k)
\end{equation}

\n gives $\arg a_1(k)$ (the symbol $\mathrm{P}$ means the principal
value):

\begin{equation}\label{eqi12}
 \arg a_1(k)=-\frac{1}{\pi}\mathrm{P}\int _{-\infty}^{\infty}\frac{\ln
 |a(k')|}{k'-k}dk'
\end{equation}

\n Therefore, from (\ref{eqi11}), (\ref{eqi12}), for real $k$:

\begin{equation}\label{eqi13}
 \ln a_1(k)= \ln |a(k)|- \frac{i}{\pi}\mathrm{P}\int _{-\infty}^{\infty}\frac{\ln
 |a(k')|}{k'-k}dk'
\end{equation}

\n With the help of the Sohotski-Plemelj formula we have (cf.
\cite{ZMNP,J99})

\begin{equation}\label{eqi15}
 \ln a_1(k)= \frac{1}{\pi i}\int _{-\infty}^{\infty}\frac{\ln
 |a(k')|}{k'-k-i0}dk',
\end{equation}

\n or with (\ref{eqi10})

\begin{equation}\label{eqi16}
 \ln a(k)=-i\alpha k +\sum
 _{n=1}^{N}\ln\frac{k-i\kappa_n}{k+i\kappa_n}+
 \frac{1}{\pi i}\int _{-\infty}^{\infty}\frac{\ln
 |a(k')|}{k'-k-i0}dk'.
\end{equation}

We will argue that (\ref{eqi15}), (\ref{eqi16}) are valid not only
for real $k$, but also when $k$ is in the upper half plane. Indeed,
from the Cauchy theorem (the closed contour $\Gamma$  consists from
the real axis and the infinite semicircle in the upper half plane,
where $\ln a_1(k)=0$) for the function $\ln a_1(k)$ and
$\mathrm{Im}\phantom{*} k>0$ we have:

\begin{equation}\label{eqi17}
 \ln a_1(k)= \frac{1}{2\pi i}\int _{-\infty}^{\infty}\frac{\ln
 a_1(k')}{k'-k}dk'.
\end{equation}

The substitution of (\ref{eqi15}) into (\ref{eqi17}) gives

\begin{equation}\label{eqi18}
 \ln a_1(k)= \frac{1}{2(\pi i)^2}\int _{-\infty}^{\infty}\Big(\int _{-\infty}^{\infty}\frac{dk'}{(k'-k)(k''-k'-i0)}\Big) \ln
 |a(k'')|dk''
\end{equation}

\n Computing the integral in the brackets with the residue theorem,
the contour $\Gamma$ being as before (note that the pole at
$k'=k''-i0$ is outside the contour, since $k''$ is real) we find

\begin{equation}\label{eqi20}
 \ln a_1(k)= \frac{1}{\pi i}\int _{-\infty}^{\infty}\frac{\ln
 |a(k'')|dk''}{k''-k},\qquad \mathrm{Im}\phantom{*} k>0.
\end{equation}

\n Therefore, from (\ref{eqi10}) and (\ref{eqi20}) when $k$ is in
the upper half plane:

\begin{equation}\label{eqi21}
 \ln a(k)=-i\alpha k +\sum
 _{n=1}^{N}\ln\frac{k-i\kappa_n}{k+i\kappa_n}+
 \frac{1}{\pi i}\int _{-\infty}^{\infty}\frac{\ln |a(k')|}{k'-k}dk'
\end{equation}

Equation (\ref{eqi3}) can also be written in the form

\begin{equation}\label{eqi22}
 \chi_x (x,k)+(\chi-ik)^2=\frac{1}{4}+\lambda (k) (m(x)+\omega)
\end{equation}

\n and admits a solution with the asymptotic expansion
\begin{equation}\label{eqi23}
 \chi(x,k)= \frac{1}{2}+i k+\sum_{n=1}^{\infty}{p_{n}}{\lambda(k)^n}.
\end{equation}

\n Since $\lambda(i/2)=0$, then $\chi(x,i/2)=0$ and therefore, due
to (\ref{eqi2})  $\ln a(i/2)=0$. Now (\ref{eqi21}) for $k=i/2$ gives
the integral $\alpha$ (\ref{eqi8}) in terms of the scattering data:

\begin{equation}\label{eqi24}
 \alpha = \sum
 _{n=1}^{N}\ln\Big(\frac{1+2\kappa_n}{1-2\kappa_n}\Big)^2-
 \frac{8}{\pi }\int _{0}^{\infty}\frac{\ln |a(\widetilde{k})|}{4\widetilde{k}^2+1}d\widetilde{k}.
\end{equation}

\n The integral $\sqrt{\omega} \alpha$ is equal to $H_{-1}$, cf.
(\ref{eq2aa}), \cite{L05,CL05,I05}:

\b \nonumber
 H_{-1}&\equiv& \int _{-\infty}^{\infty}\Big(\sqrt{\omega+m(x)}-\sqrt{\omega}\Big)dx \phantom{********}\\
 \nonumber &=&\sqrt{\omega}\sum
 _{n=1}^{N}\ln\Big(\frac{1+2\kappa_n}{1-2\kappa_n}\Big)^2-
 \frac{8\sqrt{\omega}}{\pi }\int _{0}^{\infty}\frac{\ln |a(\widetilde{k})|}{4\widetilde{k}^2+1}d\widetilde{k}.
\e

From (\ref{eq13a}) we know that $|\mathcal{R}(k)|$ determines
$|a(k)|$. But  $|a(k)|$ determines uniquely $a(k)$ for real $k$ due
to (\ref{eqi16}) and (\ref{eqi24}). Therefore $|\mathcal{R}(k)|$
determines uniquely $a(k)$ for real $k$. Since $b(k)$ is simply
$a(k) \mathcal{R}(k)$, then as expected, the entire information
about $T(k)$  (\ref{T}) is provided by $\mathcal{R}(k)$ for $k>0$.

 The integrals $I_{-n}$ can be expressed by the scattering data
from (\ref{eqi7}) and (\ref{eqi21}) taking the expansion at
$|k|\rightarrow\infty$ : $I_{-2n}=0$;

\begin{equation}\label{eqi25}
 I_{-(2n+1)}= \frac{2i(-1)^{n+1}}{2n+1}\sum
 _{n=1}^{N}\kappa_n^{2n+1} +\frac{2i}{\pi }\int _{0}^{\infty}\widetilde{k}^{2n}\ln |a(\widetilde{k})|d\widetilde{k}.
\end{equation}

For example, from $I_{-1}$, expressed from (\ref{eqi8aaa}),
(\ref{eqi25}) and using (\ref{eqi24}), the conservation law $H_{-2}$
(see (\ref{eq2aa})) can be expressed by the scattering data:

\b\label{eqi26}
 H_{-2}\equiv -\frac{1}{4}\int _{-\infty}^{\infty}\Big(\frac{1}{\sqrt{q}}-\frac{1}{\sqrt{\omega}}
 +\frac{q_{x}^{2}}{4q^{5/2}}\Big)dx= \phantom{*******************}\nonumber
 \\ \nonumber
 -\frac{1}{4\sqrt{\omega}}\sum_{n=1}^{N}\Big(\ln\Big(\frac{1+2\kappa_n}{1-2\kappa_n}\Big)^2-16\kappa_n\Big)
 -\frac{2}{\pi\sqrt{\omega} }\int _{0}^{\infty}\frac{8\widetilde{k}^2+1}{4\widetilde{k}^2+1}\ln |a(\widetilde{k})|d\widetilde{k}.
\e

From (\ref{eqi22}) and (\ref{eqi23}) we obtain (recall that
$q=m+\omega=u-u_{xx}+\omega$),

\b \label{eqi27} p_1+p'_{1}=q, \qquad  p_{1}=u-u_{x}+\omega, \e

\n which leads to the integral:

\b \label{eqi28} H_{0}=\int _{-\infty}^{\infty} m dx, \e

\n i.e.

\b \label{eqi29}\int _{-\infty}^{\infty} p_{1} dx=H_0+ \int
_{-\infty}^{\infty} \omega dx \e

\n The infinite contribution $\int _{-\infty}^{\infty} dx$ does not
make  sense, but all such contributions cancel the also infinite
term $\int _{-\infty}^{\infty} (1/2+i k) dx$ from (\ref{eqi23}) when
it is substituted in (\ref{eqi2}).

The next equation from (\ref{eqi22}) and (\ref{eqi23}) is

\b \label{eqi30} p_{2}+p'_{2}+p_{1}^{2}=0, \e

\n and hence, formally (recall (\ref{eq2a}), (\ref{eqi28}))

\b \label{eqi31} \int _{-\infty}^{\infty} p_{2}dx=-\int
_{-\infty}^{\infty} p_{1}^{2}dx=-2H_1 - 2\omega H_{0}-\int
_{-\infty}^{\infty} \omega^2 dx . \e

\n From (\ref{eqi22}) and (\ref{eqi23}) the equation for $p_{3}$ is

\b \label{eqi32} p_{3}+p'_{3}+2p_{1}p_{2}=0, \e

\n and the next conserved quantity $\label{eqi33} \int
_{-\infty}^{\infty}p_{3} dx = -2 \int _{-\infty}^{\infty}
p_{1}p_{2}dx$ is  (using (\ref{eqi27}), (\ref{eqi30}), (\ref{eq2b})
-- see the technicalities described in \cite{I05})

\b \label{eqi35}  \int _{-\infty}^{\infty} p_3dx =4H_2 +4\omega
H_{1} +6\omega ^{2} H_{0}+2\int _{-\infty}^{\infty} \omega^3 dx . \e

Now, let us expand $\ln a(k)$ about the point $k=i/2$. To this end,
for simplicity, we define a new parameter, $l$, such that:

\b \label{eqi36}  k\equiv\frac{i}{2}(1+4l)^{1/2} , \qquad
\lambda\equiv \frac{l}{\omega}. \e

\n The expansion is now about $l=0$. Using (\ref{eqi36}) and
(\ref{eqi23}) in (\ref{eqi2}), and the expressions (\ref{eqi29}),
(\ref{eqi31}), (\ref{eqi35}) we finally obtain

\b \label{eqi37}  \ln
a(k(l))=\frac{l}{\omega}H_0-\Big(\frac{l}{\omega}\Big)^2(2H_1+2\omega
H_0)+\Big(\frac{l}{\omega}\Big)^3(4H_2+4\omega H_1+6\omega^2
H_0)\nonumber
\\ +o(l^3).\phantom{**} \e

Now the expansion of (\ref{eqi21}) in $l$ (\ref{eqi36}), taking into
account (\ref{eqi24}), (\ref{eqi37}) gives the Hamiltonians in terms
of the scattering data:

\b \label{eqi38} H_0=2\omega \sum
 _{n=1}^{N}\Big(\ln \frac{1+2\kappa_n}{1-2\kappa_n}+\frac{4\kappa _n}{1-4\kappa_n
 ^{2}}\Big)-
 \frac{16\omega}{\pi }\int _{0}^{\infty}\frac{\ln |a(\widetilde{k})|}{(4\widetilde{k}^2+1)^2}d\widetilde{k}, \e

\b \label{eqi39} H_1=\omega^2 \sum
 _{n=1}^{N}\Big(\ln \frac{1-2\kappa_n}{1+2\kappa_n}+\frac{4\kappa _n(1+4\kappa_n
 ^{2})}{(1-4\kappa_n ^{2})^2}\Big)+ \frac{128\omega^2}{\pi }\int _{0}^{\infty}
 \frac{\widetilde{k}^2 \ln |a(\widetilde{k})|}{(4\widetilde{k}^2+1)^3}d\widetilde{k}, \nonumber \\ \e

\b  H_2=\omega^3 \sum
 _{n=1}^{N}\Big(\ln \frac{1-2\kappa_n}{1+2\kappa_n}+\frac{4\kappa _n(3+32\kappa_n
 ^{2}-48\kappa _n ^4)}{3(1-4\kappa_n ^{2})^3}\Big)\phantom{**********}\nonumber \\ \label{eqi40} -\frac{8\omega^3}{\pi }\int _{0}^{\infty}
 \frac{(-5+28\widetilde{k}^2 +80\widetilde{k}^4+64\widetilde{k}^6)\ln |a(\widetilde{k})|}{(4\widetilde{k}^2+1)^4}d\widetilde{k}. \e

 \n In the same fashion the higher conservation laws (which are
 nonlocal) can be expressed through the scattering data.

\section{Poisson brackets of the scattering data} \label{PB}

In this section our aim will be to compute the Poisson brackets
between the elements of the scattering matrix (\ref{T}). Let us
consider, for example, $\{a(k_1),b(k_2)\}$:
\begin{equation}\label{20}
 \{a(k_{1}),b(k_{2})\}=-\int_{-\infty}^{\infty} q(x)\Big(\frac{\delta a(k_{1})}{\delta m(x)}\frac{\partial}{\partial x}
 \frac{\delta b(k_{2})}{\delta m(x)}-\frac{\delta b(k_{2})}{\delta m(x)}\frac{\partial}{\partial x} \frac{\delta a(k_{1})}{\delta
 m(x)}\Big)dx.
\end{equation}

\n For the computation of $\delta a(k )/\delta m(x)$ and $\delta
b(k)/\delta m(x)$ we use (\ref{eq10a}) and (\ref{eq10b}):

\b \frac{\delta a(k)}{\delta m(x)}=(2ik)^{-1} \Big(\frac{\delta
\varphi(y,k)}{\delta m(x)}\frac{\partial}{\partial
y}\bar{\psi}(y,k)-\frac{\delta \bar{\psi}(y,k)}{\delta
m(x)}\frac{\partial}{\partial y}\varphi(y,k)
 \nonumber \\\label{eq21}
+\varphi(y,k)\frac{\partial}{\partial y}\frac{\delta
\bar{\psi}(y,k)}{\delta
m(x)}-\bar{\psi}(y,k)\frac{\partial}{\partial y}\frac{\delta
\varphi(y,k)}{\delta m(x)}\Big).\e

\n The function $G(x,y,k)\equiv \delta \varphi(y,k)/\delta m(x)$
satisfies the equation, obtained as a variational derivative of
(\ref{eq3}):

\b \label{eq22} \Big(\partial^{2}_{y}-\lambda
m(y)+k^{2}\Big)G(x,y,k)=\lambda \delta(x-y) \varphi(y,k).\e

\n Since the source on the right hand side of (\ref{eq22}) is zero
for $y<x$ (due to the delta-function) and since $\varphi(y,k)$ is
defined by its asymptotic when $y\rightarrow-\infty$, i.e. for
$y\rightarrow-\infty$, $\varphi(y,k)$ does not depend on $m(x)$, the
solution of (\ref{eq22}) must satisfy

\b \label{eq23} G(x,y,k)=0, \qquad y<x.\e

\n $G(x,y,k)$, considered as a solution of (\ref{eq22}) is a
continuous function of $y$ for all $y$, however due to the source on
the right hand side (proportional to a delta function)  $\partial
G(x,y,k)/\partial y$ has a finite jump at $y=x$. The value of
$\partial G(x,y,k)/\partial y$ for $y\rightarrow x+0$ can be found
by integrating both sides of (\ref{eq22}) from $x-\varepsilon$ to
$x+\varepsilon$ and then taking $\varepsilon\rightarrow +0$:

\b \label{eq24} \frac{\partial G(x,y,k)}{\partial y}\Big
|_{y=x+0}=\lambda \varphi(x,k).\e

Now we can make use of the fact, that the left hand side of
(\ref{eq21}) does not depend on $y$. We take $y=x+\varepsilon$,
$\varepsilon>0$ and then we take $\varepsilon\rightarrow0$. Since
$\psi(y,k)$ is defined by its asymptotic when $y\rightarrow\infty$,
i.e. for $y\rightarrow\infty$, $\psi(y,k)$ does not depend on
$m(x)$, by an analogous arguments we conclude that  $\delta
\psi(y,k)/\delta m(x)=0$ for $y>x$. Then finally from (\ref{eq21})
it follows:

\b\label{eq25} \frac{\delta a(k)}{\delta m(x)}=-\frac{\lambda}{2ik}
\bar{\psi}(x,k)\varphi(x,k) .\e

Similarly, we find

\b\label{eq26} \frac{\delta b(k)}{\delta m(x)}=\frac{\lambda}{2ik}
\psi(x,k)\varphi(x,k) .\e

Substituting (\ref{eq25}), (\ref{eq26}) in (\ref{eq21}), we have

\b \{a(k_{1}),b(k_{2})\}= \frac{\lambda (k_{1})\lambda
(k_{2})}{(2i)^{2}k_{1}k_{2}}\times \phantom{***********************}
\nonumber
\\ \int_{-\infty}^{\infty} q(x)\Big(\bar{\psi}(x,k_{1})\varphi(x,k_{1})\frac{\partial}{\partial
x}(\psi(x,k_{2})\varphi(x,k_{2})) \phantom{*****}
 \nonumber \\\label{eq27}
-\psi(x,k_{2})\varphi(x,k_{2})\frac{\partial}{\partial
x}(\bar{\psi}(x,k_{1})\varphi(x,k_{1}))\Big)dx.\e

The expression under the integral in (\ref{eq27}) is a total
derivative. Indeed, let $f_{1}$, $g_{1}$, $f_{2}$, $g_{2}$ be two
pairs of solutions of (\ref{eq3}) with spectral parameters $k_{1}$
and $k_{2}$ correspondingly, i.e.

\b  \partial^{2}_{x} f_{1,2}=\Big(\frac{1}{4}+\lambda(k_{1,2})
q(x)\Big)f_{1,2}, \nonumber \\ \label{eq28} \partial^{2}_{x}
g_{1,2}=\Big(\frac{1}{4}+\lambda(k_{1,2}) q(x)\Big)g_{1,2}.\e

\n Then with (\ref{eq28}) one can check easily the following
identity: \b q(x)\Big(f_{1}g_{1}\frac{\partial}{\partial
x}(f_2g_2)-f_{2}g_{2}\frac{\partial}{\partial x}(f_1g_1)\Big)=
\phantom{*************} \nonumber
\\ \frac{1}{\lambda (k_{2})-\lambda (k_{1})}\Big((g_{1}\partial_{x}g_{2}-g_{2}\partial_{x}g_{1})
(f_{1}\partial_{x}f_{2}-f_{2}\partial_{x}f_{1})\Big)_{x}.\label{eq31}
\e

\n Now we can take $f_1=\bar{\psi}(x,k_1)$, $f_2=\psi(x,k_2)$,
$g_1=\varphi(x,k_1)$, $g_2=\varphi(x,k_2)$ and substitute in
(\ref{eq27}). Then with (\ref{eq31}) and with the asymptotic
representations for $x\rightarrow\infty$

\b \label{eq32}\psi(x,k)\rightarrow e^{-ikx}, \qquad
\varphi(x,k)\rightarrow a(k)e^{-ikx}+b(k)e^{ikx},\e and for
$x\rightarrow-\infty$

\b \label{eq33}\varphi(x,k)\rightarrow e^{-ikx}, \qquad
\psi(x,k)\rightarrow \bar{a}(k)e^{-ikx}-b(k)e^{ikx}\e we obtain

\b \{a(k_{1}),b(k_{2})\}= \frac{\lambda (k_{1})\lambda
(k_{2})}{(2i)^{2}k_{1}k_{2}(\lambda (k_2)-\lambda (k_1))}\times
\phantom{************} \nonumber
\\ \Big[\lim_{x\rightarrow\infty}
\Big((k_{1}^{2}-k_{2}^{2})a(k_1)a(k_2)e^{-2ik_2
x}-(k_{1}^{2}-k_{2}^{2})b(k_1)b(k_2)e^{2ik_1 x} \nonumber
\\ +(k_{1}+k_{2})^{2}a(k_1)b(k_2)-(k_{1}+k_{2})^{2}b(k_1)a(k_2)e^{2i(k_1-k_2)
x}\Big)\nonumber \\ -\lim_{x\rightarrow-\infty}
\Big((k_{1}^{2}-k_{2}^{2})a(k_1)\bar{a}(k_2)e^{-2ik_2
x}-(k_{1}^{2}-k_{2}^{2})\bar{b}(k_1)b(k_2)e^{-2ik_1 x}
 \nonumber \\\label{eq34}
-(k_1-k_2)^{2}a(k_1)b(k_2)+(k_1-k_2)^{2}\bar{b}(k_1)\bar{a}(k_2)e^{-2i(k_1+k_2)x}\Big)\Big].\e

\n The expression on the right hand side in (\ref{eq34}) is defined
only as a distribution. Then applying the formula
$\lim_{x\rightarrow\infty}\mathrm{P}\frac{e^{ikx}}{k}=\pi i \delta
(k)$ and assuming $k_{1,2}>0$ we have

\b   \{\ln a(k_{1}), \ln b(k_{2})\}= \omega \lambda (k_{1})\lambda
(k_{2})\Big(-\frac{k_{1}^{2}+k_{2}^{2}}{2k_1k_2(k_{1}^{2}-k_{2}^{2})}+\frac{
\pi i}{2k_1} \delta(k_1-k_2) \Big).\nonumber \\\e

In the same fashion the rest of the Poisson brackets between the
scattering data can be computed. The result can be expressed in
terms of the quantities \b \label{eq36} \rho(k)\equiv-\frac{2k}{\pi
\omega \lambda(k)^2}\ln|a(k)|,\qquad \phi(k)\equiv \arg b(k), \qquad
k>0:\e

\n Their Poisson brackets have the canonical form

\b \label{eq37} \{\phi(k_1),\rho(k_2)\}=\delta(k_1-k_2),\quad
\{\phi(k_1),\phi(k_2)\}=\{\rho(k_1),\rho(k_2)\}=0\e

\n and thus (\ref{eq36}) are the action-angle variables for the CH
equation, related to the continuous spectrum. Note that from
(\ref{eq37}) and (\ref{eqi39}) we have

\b \label{eq38} \dot{\phi}=\{\phi,H_1\}= \frac{k}{\lambda(k)},\e

\n which agrees with (\ref{eq18}).

Let us now concentrate on the discrete spectrum. Let us denote, for
simplicity, $\lambda_n\equiv\lambda(i\kappa_n)$.  We will need the
variational derivatives $\delta \lambda_n/\delta m(x)$ and $\delta
b_n/\delta m(x)$. Due to (\ref{eq202}), for $\delta b_n/\delta m(x)$
the expression (\ref{eq26}) will be formally used followed by taking
the limit $k\rightarrow i\kappa_n$. In order to find $\delta
\lambda_n/\delta m(x)$ we proceed as follows. Differentiating the
equation

\b \label{eq39} \varphi^{(n)}_{xx}=\Big(\frac{1}{4}+\lambda_n
q(x)\Big)\varphi^{(n)},\e

\n we obtain ($\delta q=\delta m$):

\b \label{eq40} \delta \varphi^{(n)}_{xx}=(\delta\lambda_n)q
\varphi^{(n)}+\lambda_n(\delta
m)\varphi^{(n)}+\Big(\frac{1}{4}+\lambda_n q\Big)\delta
\varphi^{(n)}.\e

From (\ref{eq39}) and (\ref{eq40}) it follows

\b \label{eq41} (\varphi^{(n)}\delta
\varphi^{(n)}_{x}-\varphi^{(n)}_{x}\delta
\varphi^{(n)})_x=(\delta\lambda_n)q[\varphi^{(n)}]^2+\lambda_n(\delta
m)[\varphi^{(n)}]^2.\e

\n The integration of (\ref{eq41}) gives:

\b \label{eq42} (\delta\lambda_n)\int
_{-\infty}^{\infty}q(x)[\varphi^{(n)}(x)]^2dx=-\lambda_n\int
_{-\infty}^{\infty}(\delta m(x))[\varphi^{(n)}(x)]^2dx.\e

\n or

\b \label{eq42a} \frac{\delta\ln \lambda_n}{\delta
m(x)}=-\frac{[\varphi^{(n)}(x)]^2}{\int
_{-\infty}^{\infty}q(y)[\varphi^{(n)}(y)]^2dy}.\e

\n From (\ref{eq3}) it is not difficult to obtain

\b \label{eq43} (\varphi
\varphi_{x\lambda}-\varphi_{x}\varphi_{\lambda})_x=q\varphi^2.\e

\n We will integrate (\ref{eq43}) and then take the limit
$k\rightarrow i\kappa_n$, i.e. $\lambda\rightarrow\lambda_n$. We
take into account that with this limit, clearly

\b \label{eq44} \varphi(x,k)&\rightarrow&b_n e^{-\kappa_n
x}+o(e^{-\kappa_n x}), \qquad x\rightarrow \infty;
 \\\label{eq45}
\varphi_{\lambda}(x,k)&\rightarrow&
\frac{a'(i\kappa_n)}{\lambda'(i\kappa_n)} e^{\kappa_n
x}+o(e^{\kappa_n x}), \qquad x\rightarrow \infty. \e

\n Therefore

\b \label{eq46} i\omega b_n a'(i\kappa_n)=\int
_{-\infty}^{\infty}q(y)[\varphi^{(n)}(y)]^2dy \e

\n and finally

\b \label{eq42b} \frac{\delta\ln \lambda_n}{\delta
m(x)}=\frac{i[\varphi^{(n)}(x)]^2}{\omega b_n a'(i\kappa_n)}.\e

\n We compute the expression

\b \{\ln \lambda _n,b_l\}= \frac{i\lambda _l}{2\omega b_n\kappa_l
a'(i\kappa_n)}\times \phantom{************************}\nonumber \\
\int_{-\infty}^{\infty} q(x) \lim _{k_j\rightarrow i\kappa_j}
\Big(\varphi^2(x,k_{n})(\psi(x,k_{l})\varphi(x,k_{l}))_x\phantom{*******}\nonumber
\\\label{eq43a} -\psi(x,k_{l})\varphi(x,k_{l})(\varphi^2(x,k_{n}))_x\Big) dx.\phantom{**}\e

\n Taking $f_1=g_1=\varphi(x,k_n)$, $f_2=\psi(x,k_l)$,
$g_2=\varphi(x,k_l)$ in (\ref{eq31}) and using the asymptotic
representations (\ref{eq32}), (\ref{eq33}) for $x\rightarrow \pm
\infty$ we obtain

\b \{\ln \lambda _n,b_l\}= \frac{i\lambda _l}{2 b_n\kappa_l
a'(i\kappa_n)}\times \phantom{**********************}\nonumber
\\\label{eq44a} \lim_{x\rightarrow \infty}\lim _{k_j\rightarrow
i\kappa_j}
\frac{(k_n+k_l)\Big(a(k_n)b(k_n)b(k_l)-a(k_l)b^2(k_n)e^{2i(k_n-k_l)x}\Big)}{k_n-k_l}.\e

\n Clearly, the right hand side of (\ref{eq44a}) is zero if
$\kappa_n\neq\kappa_l$, since $a(i\kappa_n)=0$. However, if
$\kappa_n=\kappa_l$, the l'Hospital's rule for the limit $\kappa_l
\rightarrow\kappa_n$ gives

\b \{\ln \lambda _n,b_l\}= -\lambda _n b_n\delta _{nl}.\e

\n If we define the quantities

\b \label{eq45a} \rho_n \equiv \lambda_n^{-1},\qquad \phi_n \equiv
-\ln |b_n|, \qquad n=1,2,\ldots,N, \e

\n their Poisson brackets have the canonical form

\b \label{eq47} \{\phi_n,\rho_l\}=\delta_{nl},\qquad
\{\phi_n,\phi_l\}=\{\rho_n,\rho_l\}=0\e

\n and thus (\ref{eq47}) are the action-angle variables for the CH
equation, related to the discrete spectrum. They also commute with
the variables on the continuous spectrum (\ref{eq36}). Note that
from (\ref{eq47}) and (\ref{eqi39}) we have

\b \label{eq48} \dot{\phi}_n=\{\phi_n,H_1\}=
\{\phi_n,\kappa_n\}\frac{\partial H_1}{\partial
\kappa_n}=-\frac{4\omega\kappa_n}{1-4\kappa_n^2},\e

\n which agrees with (\ref{eq207}).

\section{Conclusions} \label{sec:1}
In this paper the action-angle variables for the CH equation are
computed. They are expressed in terms of the scattering data for
this integrable system, when the solutions are confined to be
functions in the Schwartz class. The important question about the
behavior of the scattering data at $k=0$ deserves further
investigation. It is possible that in the case of singular behavior
the Poisson bracket has to be modified in a way, similar to the KdV
case, as described in \cite{FT85,APP88,BFT86}. The case $\omega=0$
(in which the spectrum is only discrete, cf. \cite{CM99}) is
presented in \cite{V05}. The situation when the condition
$m(x,0)+\omega>0$ does not hold requires separate analysis
\cite{K05,B04} (if $m(x,0)+\omega$ changes sign there are infinitely
many positive eigenvalues accumulating at infinity,  cf.
\cite{C01}). The action-angle variables for the periodic CH equation
and $\omega=0$ are reported in \cite{P05}.

\section*{Acknowledgements} The authors acknowledge the support of the
Science Foundation Ireland, Grant 04/BR6/M0042.



\end{document}